\documentclass[english]{article}
\usepackage[T1]{fontenc}
\usepackage[latin9]{inputenc}
\usepackage{geometry}
\usepackage{color}
\usepackage{array}

\makeatletter

\providecommand{\tabularnewline}{\\}

\makeatother

\usepackage{babel}

\begin{document}

\title{Efficient protocols for unidirectional and bidirectional controlled
deterministic secure quantum communication: Different alternative
approaches}

\author{Anirban Pathak}

\maketitle
\begin{center}
Jaypee Institute of Information Technology, A-10, Sector-62, Noida,
Up-201307, India
\par\end{center}
\begin{abstract}
Recently, Hassanpour and Houshmand have proposed a protocol of controlled
deterministic secure quantum communication (Quant. Info. Process,
DOI 10.1007/s11128-014-0866-z (2014)). The authors compared the efficiency
of their protocol with that of two other existing protocols and claimed
that their protocol is efficient. Here, we have shown that the efficiency
of Hassanpour Houshmand (HH) protocol is not high, and there exist
several approaches through which more efficient protocols for the
same task can be designed. To establish this point, we have proposed
an efficient protocol of controlled deterministic secure quantum communication
which is based on permutation of particles (PoP) technique and is
considerably efficient compared to HH protocol. We have also generalized
this protocol into its bidirectional counterpart. Interestingly, bipartite
entanglement (Bell state) is sufficient for the realization of the
proposed protocols, but HH protocol and other existing protocols require
at least tripartite entanglement. Further, we have shown that it is
possible to construct a large number of efficient protocols of unidirectional
and bidirectional controlled deterministic secure quantum communication
by using various alternative approaches and different quantum states.
These alternative protocols can be realized by modifying the existing
protocols of quantum secure direct communication and deterministic
secure quantum communication. We have also shown that it is possible
to design completely orthogonal-state-based protocols for unidirectional
and bidirectional controlled deterministic secure quantum communication.
\end{abstract}

\section{Introduction}

In 1984, Bennett and Brassard proposed a protocol of quantum key distribution
(QKD) \cite{bb84}. The protocol, which is popularly known as BB84
protocol, drew considerable attention of the cryptographic community
as the protocol is unconditionally secure. As a consequence, several
other protocols of secure quantum communication have been proposed.
Interestingly, in the early years of secure quantum communication,
only protocols of QKD were proposed \cite{bb84,ekert,b92,vaidman-goldenberg}.
However, it was realized soon that quantum states can be employed
to design the protocols for secure direct quantum communication where
we can circumvent the prior generation of keys (i.e., QKD), and thus,
directly communicate a message by using quantum resources. In the
last few years, many such protocols of secure direct quantum communication
have been proposed. Such protocols can be classified into two broad
classes: (a) protocols for quantum secure direct communication (QSDC)
\cite{Long   and   Liu,ping-pong,for PP,lm05} and (b) protocols for
\textit{\emph{deterministic secure quantum communication}}\emph{ }(DSQC)
\cite{dsqc_summation,dsqcqithout-max-entanglement,dsqcwithteleporta,entanglement      swapping,Hwang-Hwang-Tsai,reordering1,the:cao and song,the:high-capacity-wstate}.
The difference between DSQC and QSDC is very small. Specifically,
in a DSQC protocol Bob (receiver) can decode the secret message sent
by Alice (sender) only after the receipt of the additional classical
information of at least one bit for each qubit transmitted by Alice.
In contrast, no such additional classical information is required
in QSDC (\cite{With Anindita-pla} and references therein). Further,
extending the idea of secure direct quantum communication, a few three-party
protocols have recently been introduced \cite{CQSDC-Hassanpour,CQSDC-Gao,CQSDC-Dong}.
In these protocols, Alice can directly communicate a secret message
to Bob, if a controller (Charlie) allows them to do so. These protocols
are referred to as controlled QSDC (CQSDC) protocols. However, if
we stick to the definition of QSDC, it does not appear justified to
refer to these protocols as CQSDC. This is so because in all these
protocols, Bob can read the message sent by Alice, only after Charlie
provides him some additional classical information (usually outcome
of a measurement performed by Charlie). Thus, it would be more appropriate
to call them controlled DSQC (CDSQC) or controlled secure direct quantum
communication. In what follows, we refer to these protocols as CDSQC.
Very recently, Hassanpour and Houshmand (HH) have proposed an interesting
entanglement swapping based protocol of CDSQC \cite{CQSDC-Hassanpour}
using GHZ-like states. They claimed their protocol as a protocol of
CQSDC, but the claim is not correct as in their protocol the receiver
(Bob) can decode the secret message only after the receipt of classical
information from both the sender (Alice) and controller (Charlie).
Thus, the HH protocol is actually a protocol of DSQC. Further, they
claimed their protocol as an efficient one by comparing the qubit
efficiency of their protocol with that of Gao et al. \cite{CQSDC-Gao}
and Dong et al. \cite{CQSDC-Dong} protocols. However, a critical
analysis of HH protocol reveals that the efficiency reported in Ref.
\cite{CQSDC-Hassanpour} can be considerably improved, and the protocols
of CDSQC can be designed using various quantum states and various
alternative ways. This paper aims to establish these facts. Further,
in Ref. \cite{CQSDC-Hassanpour}, Hassanpour and Houshmand have mentioned
that in future they wish to extend the HH protocol to a protocol of
controlled bidirectional deterministic secure quantum communication
(CBDSQC). Here, we explicitly provide a protocol of CBDSQC and also
provide some possible alternative methods for realization of CBDSQC.
In our protocols, PoP plays a very important role. This technique
was first introduced by Deng and Long in 2003, while they proposed
a protocol of QKD based on this technique \cite{PoP}. In what follows,
we will find that PoP helps us to improve the efficiency of the protocols
of CDSQC and CBDSQC and  to obtain completely orthogonal-state-based
protocols of CDSQC and CBDSQC.

Remaining part of this paper is organized as follows. In Section \ref{sec:Controlled-secure-direct},
we provide a protocol of CDSQC that uses particle order permutation
(PoP), dense coding and Bell states. Subsequently, we show that there
exist several alternative approaches through which CDSQC protocols
can be designed. In Section \ref{sec:Controlled-bidirectional-secure},
we show that our main protocol of CDSQC can be turned to a protocol
of CBDSQC and there exist infinitely many quantum channels that can
be used to realize CBDSQC. In Section \ref{sec:Qubit-efficiency-of},
we compared the efficiency of the proposed protocols with that of
the HH protocol and established that the efficiency of CDSQC protocol
can be increased considerably. To be precise, we have shown that our
protocols are much more efficient compared to the HH protocol. Finally,
we conclude the paper in Section \ref{sec:Conclusions}, where we
categorically list a set of important observations.

\section{Controlled secure direct quantum communication using Bell states\label{sec:Controlled-secure-direct}}

Before we elaborate our protocol of CDSQC, it will be useful to briefly
describe the basic ideas of some closely connected protocols of QSDC
and DSQC. Let us first describe the Ping-Pong (PP) protocol \cite{ping-pong}
of QSDC. In PP protocol, Bob prepares a Bell state, keeps the second
qubit with himself and sends the first qubit to Alice, who encodes
a message on the qubit she received and returns the qubit to Bob.
Subsequently, Bob performs a joint measurement on both the qubits
using Bell basis and obtains the secret encoded by Alice. The encoding
is done by following a pre-decided rule: to communicate 0, Alice does
nothing and to communicate 1, she applies $X$ gate. Thus, if Bob
prepares $|\psi^{+}\rangle=\frac{|00\rangle+|11\rangle}{\sqrt{2}}$
as the initial state and receives the same state as the outcome of
his measurement then Alice has encoded 0, whereas if his measurement
yields $|\phi^{+}\rangle=\frac{|10\rangle+|01\rangle}{\sqrt{2}}$
then Alice has encoded 1. Clearly, in PP protocol, full advantage
of dense coding is not utilized. The same is utilized in Cai Li (CL)
protocol \cite{cai and li PRA}, which is a modified PP protocol with
the following rule for encoding: to encode $00,\,01,\,10$ and $11$,
Alice applies $I,\, X,\, iY$ and $Z$ gates respectively. Now, we
can easily see that Bob's possible measurement outcomes $|\psi^{+}\rangle,\,|\phi^{+}\rangle,\,|\phi^{-}\rangle,$
and $\,|\psi^{-}\rangle$ correspond to Alice's secret bits $00,\,01,\,10,$
and $11$, respectively. Except this difference in the encoding part
the CL protocol is same as the PP protocol. In both of these protocols,
Bob prepares the maximally entangled state and sends a part of it
to Alice. In contrast to PP and CL protocols, in Deng Long Liu (DLL)
protocol \cite{DLL}, Alice prepares a Bell state and sends the second
qubit to Bob and verifies that he has received it without eavesdropping.
Subsequently, Alice encodes her message in the second qubit by following
the same encoding rule as done in the CL protocol and sends the qubit
to Bob, who measures it in Bell basis. Here, it is important to note
the following: (i) If a third party (Charlie) distributes the entanglement
to Alice and Bob, then CL and DLL protocols are equivalent (this will
be the case with the protocol proposed below). (ii) Clearly, CL and
DLL protocols are more efficient than PP protocol. Further, CL and
DLL protocols can be easily realized using other entangled states
where dense coding is possible, but the efficiency will be higher
for those states where maximal dense coding is possible. Consequently,
the efficiency of any dense coding based scheme that uses multipartite
entangled states with odd number of qubits have to be lower than the
situation where the same scheme is realized using Bell states. So
far, we have not discussed anything about the security of PP, CL and
DLL schemes. The security is achieved in two alternative ways. We
refer to these two alternative methods as BB84 subroutine and GV subroutine.
In both the methods, to transmit a sequence $A$ of $n$ message qubits,
the sender creates an additional sequence $D$ of $n$ decoy qubits
and inserts the decoy qubits randomly into the sequence of message
qubits. The combined sequence is sent to the receiver. In BB84 subroutine,
the decoy qubits are prepared in a random sequence of $\left\{ |0\rangle,|1\rangle,|+\rangle,|-\rangle\right\} $,
whereas in GV subroutine $n$ decoy qubits are prepared as $|\psi^{+}\rangle^{\otimes\frac{n}{2}}$. 
\begin{description}
\item [{BB84~subroutine:}] In this method eavesdropping is checked using
two or more mutually unbiased bases (i.e., using conjugate coding)
in a manner similar to what followed in BB84 protocol \cite{bb84}.
Specifically, after receiving the authentic acknowledgment from the
receiver, the sender announces the position of the decoy qubits. Now,
the receiver measures all the decoy qubits randomly in $\left\{ |0\rangle,\,|1\rangle\right\} $
or $\left\{ |+\rangle,\,|-\rangle\right\} $ basis and announces which
basis she/he has used to measure a particular decoy qubit, position
of that decoy qubit and outcome. The sender compares the initial states
of the decoy qubits with the outcomes of the receiver's measurement
in all those cases where the basis used by the sender to prepare the
decoy qubit is same as the basis used by the receiver to measure it.
Ideally, in the absence of eavesdropping the states should match the
outcomes, while any eavesdropping effort would lead to a mismatch.
\item [{GV~subroutine:}] After receiving the authenticated acknowledgment
from the sender that he/she has received all the $2n$ qubits sent
to him/her, the sender discloses the actual sequence of the decoy
qubits, so that the receiver can perform Bell measurement on partner
particles (original Bell pairs) and reveal any effort of eavesdropping
through the disturbance introduced by Eve's measurements. \\
To understand this point, consider a that the sender prepares
$|\psi^{+}\rangle^{\otimes2}=|\psi^{+}\psi^{+}\rangle_{1234}$ and
randomly changes the sequence of the particles and sends them to receiver.
Now, also consider that Eve knows that two Bell states are sent, but
she does not know which qubit is entangled to which qubit. Consequently,
any wrong choice of partner particles would lead to entanglement swapping
(say, if Eve does Bell measurement on qubit numbers $13$, $14$ and/or
$24$ that would lead to entanglement swapping). Now, at a later time,
when the sender discloses the actual sequence of the transmitted qubits,
the receiver uses that data to rearrange the qubits into the original
sequence and performs Bell measurement on them. Clearly, attempts
of eavesdropping will leave detectable traces through the entanglement
swapping and whenever receiver's Bell measurement would yield any
result other than $|\psi^{+}\rangle$, they will know the existence
of an eavesdropper.
\end{description}
In conventional PP, CL and DLL protocols, eavesdropping is checked
by applying BB84 subroutine (i.e., using conjugate coding), but the
encoding and decoding is done by using orthogonal states. Thus, if
we replace BB84 subroutine by GV subroutine, then we can obtain completely
orthogonal state based counterparts of PP, CL and DLL protocols. Specifically,
orthogonal-state-based protocols of QSDC that correspond to PP, CL
and DLL are referred to as ${\rm PP^{GV},\, CL^{GV}}$ and ${\rm DLL^{GV}}$,
respectively \cite{With preeti,my book}. This discussion provides
us sufficient background to describe an efficient protocol of CDSQC
which can be realized either using conjugate coding (if BB84 subroutine
is used for eavesdropping checking) or in a way which is completely
orthogonal-state-based (if GV subroutine is used for eavesdropping
checking). Our protocol can be described in following steps:
\begin{enumerate}
\item Charlie prepares $n$ Bell states $|\psi^{+}\rangle^{\otimes n}$
with $n\geq2$ . He uses the Bell states to prepare 2 ordered sequences
as follows: 

\begin{enumerate}
\item A sequence with all the first qubits of the Bell states: $P_{A}=\left[p_{1}\left(t_{A}\right),p_{2}\left(t_{A}\right),...,p_{n}\left(t_{A}\right)\right]$, 
\item A sequence with all the second qubits of the Bell states: $P_{B}=[p_{1}(t_{B}),p_{2}(t_{B}),...,p_{n}(t_{B})]$,
\end{enumerate}
where the subscripts $1,2,\cdots,n$ denote the order of a particle
pair $p_{i}=\{t_{A}^{i},t_{B}^{i}\},$ which is in the Bell state. 

\item Charlie applies an $n$-qubit permutation operator $\Pi_{n}$ on $P_{B}$
to create a new sequence as $P_{B}^{\prime}=\Pi_{n}P_{B}$ . The actual
order ($\Pi_{n}$) is known to Charlie only. 
\item Charlie prepares $2n$ decoy qubits either as a random sequence of
$\left\{ |0\rangle,|1\rangle,|+\rangle,|-\rangle\right\} $ or as
$|\psi^{+}\rangle^{\otimes n}$ and randomly inserts first (last)
$n$ decoy qubits in $P_{A}$ ($P_{B}^{\prime}$) to yield a larger
sequence $P_{A}^{\prime\prime}(P_{B}^{\prime\prime})$ having $2n$
qubits. Subsequently, Charlie sends $P_{A}^{\prime\prime}$ and $P_{B}^{\prime\prime}$
to Alice and Bob, respectively. The actual positions of the decoy
qubits are known to Charlie only. 
\item Charlie discloses the coordinates of the decoy qubits after receiving
the authenticated acknowledgment of the receipt of the qubits from
Alice and Bob. Alice and Bob apply BB84 subroutine, if the decoy qubits
are prepared as a random sequence of $\left\{ |0\rangle,|1\rangle,|+\rangle,|-\rangle\right\} $
to check the error rate (eavesdropping). Otherwise, (i.e., if the
decoy qubits are prepared as $|\psi^{+}\rangle^{\otimes n}$) they
apply a GV subroutine to check error rate. If the computed error rate
is found lower than the tolerable error limit, they go to the next
step, otherwise, they return back to Step 1.
\item On successful completion of the error checking, Alice (Bob) understands
that she (he) shares entanglement with Bob (Alice). Alice can now
encode her secret message by performing a Bell-state-based protocol
of secure direct communication (either a protocol of QSDC or that
of DSQC) and send her qubits to Bob (after randomly inserting $n$
decoy qubits). The specific QSDC/DSQC protocol may be chosen from
a large class of protocols. For example, they may use one protocol
from the following set of protocols \{${\rm PP,PP^{GV},CL,CL^{GV},DLL,DLL^{GV}}$\}.\\
Since the sequence with Alice and Bob are different, even if Alice
or Bob obtains the access of both $P_{A}$ and $P_{B}^{\prime}$,
they will not be able to find out which particle is entangled with
which particle. Thus, until Charlie discloses $\Pi_{n}$, Bob will
not be able to decode the information encoded by Alice.
\item Charlie discloses $\Pi_{n}$, when he plans to allow Bob to decode
the information encoded by Alice.
\item Since the initial Bell states and exact sequence are known, Bob measures
the initially entangled (partner) particles in Bell basis and using
the outcomes of his measurement and the knowledge of the initial state,
he decodes the secret sent by Alice. 
\end{enumerate}
Here, it is interesting to note that the above scheme is not limited
to Bell state. There exist several alternative approaches through
which efficient CBDSQC schemes can be designed. Here, we list a few
alternative paths.

\subsection{Alternative 1: CDSQC using entangled states other than Bell states\label{sub:Alternative-1:-CDSQC}}

It can be implemented by distributing other entangled states, too.
To illustrate this point more clearly, we may note that dense coding
schemes for a large set of entangled states (e.g., $W$ state, GHZ
state, GHZ-like state, $Q_{4}$ state, $Q_{5}$ state, cluster state,
$|\Omega\rangle$ state, Brown state, etc.) are described in our recent
works \cite{With chitra-ijtp,qd}. In all such dense coding schemes
using $N$-qubit entangled states, initially Alice possesses $p$
qubits (with $\mbox{\ensuremath{\frac{N}{2}\leq}}p<N)$ and Bob possesses
$(N-p)$ qubits. Subsequently, Alice encodes her message on the $p$
qubits of her possession and sends the message-encoded qubits to Bob,
who measures the $N$-qubit state in an appropriate basis. Now, to
implement a scheme of CDSQC, Charlie should follow the same strategy
as above and distribute $p$ qubits of each of the $N$-qubit entangled
states to Alice (with appropriate security measures) and a reordered
sequence of remaining $(N-p)$ qubits to Bob. Alice will encode her
secret on her qubits by applying the unitary operations that lead
to dense coding and send the qubits to Bob who will be able to decode
the message only after Charlie's disclosure of the actual order. Thus,
this provides several possibilities of obtaining CDSQC. Specifically,
CDSQC is possible using above protocol and $n$ copies of any of the
following entangled states: $W$ state, GHZ state, GHZ-like state,
$Q_{4}$ state, $Q_{5}$ state, cluster state, $|\Omega\rangle$ state,
Brown state, etc. Interestingly, possibilities are not exhausted here.
It is also possible to design entanglement swapping based protocols
of CDSQC using various other entangled states. In the following subsection,
we elaborate this point.

\subsection{Alternative 2: CDSQC using entanglement swapping and PoP \label{sub:Alternative-2:-CDSQC}}

In a recent work \cite{dsqc-ent swap}, we have shown that it is possible
to design a protocol of DSQC that can transmit an $s$-bit message
using entanglement swapping, and the quantum states of the form \begin{equation}
|\psi\rangle=\frac{1}{\sqrt{2^{s}}}\sum_{i=1}^{2^{s}}|e_{i}\rangle|f_{i}\rangle,\label{eq:state of interest}\end{equation}
where $\left\{ |e_{i}\rangle\right\} $ is a basis set in $C^{2^{m}}:\, m\geq s$
(where $|e_{i}\rangle$ is a maximally entangled $m$-qubit state)
and each of the basis vectors is an $m$-qubit maximally entangled
state ( $m\geq2)$, and $\left\{ |f_{i}\rangle\right\} $ is a basis
set in $C^{2^{l}}:l\geq s\geq1$. Elements of $\left\{ |f_{i}\rangle\right\} $
may be separable. Thus, $|\psi\rangle$ is an $m+l$ qubit state.
Further, since $\left\{ |e_{i}\rangle\right\} $ and $\left\{ |f_{i}\rangle\right\} $
are basis sets, $i\neq i^{\prime}$ implies that $|\psi\rangle$ is
an entangled state. In general, we demand $|e_{i}\rangle$ is a maximally
entangled $m$-qubit state. In our original protocol, we had assumed
that the quantum state $|\psi\rangle$ described in Eq. (\ref{eq:state of interest})
is prepared by Alice, who keeps first $m$ qubits with herself and
sends the remaining $l$ qubits to Bob in a \emph{non-clonable manner}.
By non-clonable manner, we mean that Alice sends the qubits to Bob
in such a way that Eve cannot clone the state $|f_{i}\rangle$. To
convert our DSQC protocol \cite{dsqc-ent swap} into an entanglement
swapping based protocol of CDSQC, it would be sufficient to consider
that Charlie prepares $|\psi\rangle^{\otimes n}$ and sends a sequence
of $nm$ qubits to Alice (the sequence contains first $m$ qubits
of each copy of $|\psi\rangle$), and he sends remaining $nl$ qubits
to Bob after applying $\Pi_{nl}$ on them. Subsequently, Alice will
encode her secret message by faithfully following the entanglement
swapping based DSQC protocol of ours \cite{dsqc-ent swap}, but Bob
will not be able to decode the message unless Charlie discloses $\Pi_{nl}$.
In \cite{dsqc-ent swap}, we have provided several examples of quantum
states of the form (\ref{eq:state of interest}) (cf. Table 1 of \cite{dsqc-ent swap})
which can be used to implement DSQC using entanglement swapping. The
above strategy of Charlie implies that all states of the form (\ref{eq:state of interest})
can be used to implement CDSQC using entanglement swapping. Specifically,
we can use GHZ state, GHZ-like state, cat state, cluster state, $\Omega$
state, $\chi$ state, brown state, etc. to implement CDSQC based on
entanglement swapping. Interestingly, there exists another alternative
approach through which a protocol of CDSQC can be designed.

\subsection{Alternative 3: CDSQC using dense coding and $(N+1)$-qubit entangled
state\label{sub:Alternative-3:-CDSQC}}

Consider that we have $n$ copies of the $(N+1)$-qubit quantum state
of the form \begin{equation}
|\psi\rangle=\frac{1}{\sqrt{2}}\left(|\psi_{1}\rangle_{A_{1,2,\cdots},pB_{1,2,\cdots,N-p}}|a\rangle_{C_{1}}\pm|\psi_{2}\rangle_{A_{1,2,\cdots},pB_{1,2,\cdots,N-p}}|b\rangle_{C_{1}}\right),\label{eq:the 5-qubit state-1}\end{equation}
where single qubit states $|a\rangle$ and $|b\rangle$ satisfy $\langle a|b\rangle=\delta_{a,b}$,
and $|\psi_{i}\rangle$ is an element of an $N$-qubit basis set S,
where dense coding is possible if the receiver possesses $p$ qubits
and the sender possesses $N-p$ qubits. The subscripts $A$, $B$
and $C$ indicate the qubits of Alice, Bob and Charlie respectively
and the condition \begin{equation}
|\psi_{1}\rangle\neq|\psi_{2}\rangle\label{eq:condition-1}\end{equation}
ensures that Charlie's qubit is appropriately entangled with remaining
$N$ qubits. In a protocol that uses a quantum state of this form,
Charlie prepares $n$ copies of the $(N+1)$-qubit quantum state $|\psi\rangle$
and transmits $p$ qubits of each state to Alice as a sequence $A$
and rest $N-p$ qubits to Bob as a sequence $B$. Charlie randomly
inserts decoy qubits in the sequences $A$ and $B$, but does not
alter the order of the qubits present in sequence $B.$ Here, Charlie's
control on the protocol arises from the fact that unless he measures
his qubit in $\{|a\rangle,|b\rangle\}$ basis, Bob is unaware of the
initial state on which Alice has encoded her secret message using
unitary operators. However, with the knowledge of Charlie's measurement
outcome, Bob will be able to decode the message sent by Alice\textcolor{blue}{.
}A very special case of the above described general state is a GHZ-like
state, where $|\psi_{i}\rangle\in\left\{ |\psi^{+}\rangle,|\psi^{-}\rangle,|\phi^{+}\rangle,|\phi^{-}\rangle:|\psi_{1}\rangle\neq|\psi_{2}\rangle\right\} $,
$|\psi^{\pm}\rangle=\frac{|00\rangle\pm|11\rangle}{\sqrt{2}},$ $|\phi^{\pm}\rangle=\frac{|01\rangle\pm|10\rangle}{\sqrt{2}}$$.$\textcolor{blue}{{} }

\section{Controlled bidirectional deterministic secure quantum communication
using Bell states\label{sec:Controlled-bidirectional-secure}}

In the recent paper of Hassanpour and Houshmand \cite{CQSDC-Hassanpour},
the authors have mentioned that in future they wish to extend their
proposal to a CBDSQC protocol. Here, we will show that it is a straightforward
exercise to transform our first protocol into a protocol of CBDSQC.
In fact, there exists a large number of alternatives through which
one can produce it. 
\begin{enumerate}
\item Charlie prepares $2n$ Bell states $|\psi^{+}\rangle^{\otimes2n}$
with $n\geq2$. He uses the Bell states to prepare 4 ordered sequences
as follows: 

\begin{enumerate}
\item A sequence with all the first qubits of the first $n$ Bell states:
$P_{A_{1}}=\left[p_{1}\left(t_{A}\right),p_{2}\left(t_{A}\right),...,p_{n}\left(t_{A}\right)\right]$, 
\item A sequence with all the first qubits of the last $n$ Bell states:
$P_{A_{2}}=\left[p_{n+1}\left(t_{A}\right),p_{n+2}\left(t_{A}\right),...,p_{2n}\left(t_{A}\right)\right]$,
\item A sequence with all the second qubits of the first $n$ Bell states:
$P_{B_{1}}=[p_{1}(t_{B}),p_{2}(t_{B}),...,p_{n}(t_{B})]$,
\item A sequence with all the second qubits of the last $n$ Bell states:
$P_{B_{2}}=[p_{n+1}(t_{B}),p_{n+2}(t_{B}),...,p_{2n}(t_{B})]$,
\end{enumerate}
where the subscripts $1,2,\cdots,2n$ denote the order of a particle
pair $p_{i}=\{t_{A}^{i},t_{B}^{i}\},$ which is in the Bell state. 

\item Charlie applies $n$-qubit permutation operators $\Pi_{n_{1}}$ and
$\Pi_{n_{2}}$ on $P_{A_{2}}$ and $P_{B_{1}}$ to create two new
sequences as $P_{A_{2}}^{\prime}=\Pi_{n_{1}}P_{A_{2}}$ and $P_{B_{1}}^{\prime}=\Pi_{n_{2}}P_{B_{1}}$
and sends the sequences $P_{A_{1}}$ and $P_{A_{2}}^{\prime}$ ($P_{B_{2}}$
and $P_{B_{1}}^{\prime}$) to Alice (Bob) after random insertion of
$n$ decoy qubits in each sequence. The actual order is known to Charlie
only. \\
It is pre-decided that the first (last) $n$ Bell states prepared
by Charlie are to be used for Alice to Bob (Bob to Alice) communication.
Clearly, the rest of the protocol will be analogous to the previous
one with the only difference that Alice (Bob) will encode her (his)
secret message on $P_{A_{1}}$ ($P_{B_{2}}$) and sends that to Bob
(Alice), who will be able to decode the encoded message only after
Charlie's disclosure of $\Pi_{n_{1}}(\Pi_{n_{2}})$. 
\end{enumerate}
Clearly, all the states for which dense coding is possible can be
used to implement this protocol. Now, consider a $2N+1$ qubit state
of the form \begin{equation}
|\psi\rangle=\frac{1}{\sqrt{2}}\left(|\psi_{1}\rangle_{A_{1,2,\cdots},pB_{1,2,\cdots,N-p}}|\psi_{2}\rangle_{B_{1,2,\cdots},pA_{1,2,\cdots,N-p}}|a\rangle_{C_{1}}\pm|\psi_{3}\rangle_{A_{1,2,\cdots},pB_{1,2,\cdots,N-p}}|\psi_{4}\rangle_{B_{1,2,\cdots},pA_{1,2,\cdots,N-p}}|b\rangle_{C_{1}}\right),\label{eq:the 5-qubit state-2}\end{equation}
where $|a\rangle,|b\rangle$ and $|\psi_{i}\rangle$ have the same
meaning as in the protocol of CDSQC described above as Alternative
3. The condition \begin{equation}
|\psi_{1}\rangle\neq|\psi_{3}\rangle,|\psi_{2}\rangle\neq|\psi_{4}\rangle\label{eq:condition}\end{equation}
ensures that Charlie's qubit is appropriately entangled with the remaining
qubits. By appropriately entangled we mean that unless Charlie measures
his qubit in $\{|a\rangle,|b\rangle\}$ basis and discloses the outcome.
Alice and Bob are not aware of the entangled states they share, and
consequently the receiver does not know upon the receipt of the sender's
information encoded qubits, how to decode the message. This provides
the desired control. Here, a measurement by Charlie in $\left\{ |a\rangle,|b\rangle\right\} $
basis will reduce the state into a product state of the form $|\psi_{i}\rangle\otimes|\psi_{j}\rangle$,
where first $p$ qubits of $|\psi_{i}\rangle$ and last $(N-p)$ qubits
of $|\psi_{j}\rangle$ are with Alice and the remaining qubits are
with Bob. Further, this enables Alice and Bob to use the state $|\psi_{i}\rangle$($|\psi_{j}\rangle$)
for Alice to Bob (Bob to Alice) communication using dense coding and
we can have a scheme of bidirectional controlled DSQC as the receivers
in both directions will be able to decode the message encoded by the
sender only after Charlie's disclosure of his measurement outcome.
Further, in completely different contexts, we have shown that the
5-qubit quantum states of the following form are useful for bidirectional
controlled state teleportation (BCST) \cite{bi-directional-ourpaper}
and controlled bidirectional remote state preparation (CBRSP) \cite{CBRSP}:
\begin{equation}
|\psi\rangle_{12345}=\frac{1}{\sqrt{2}}\left(|\psi_{1}\rangle_{A_{1}B_{1}}|\psi_{2}\rangle_{A_{2}B_{2}}|a\rangle_{C_{1}}\pm|\psi_{3}\rangle_{A_{1}B_{1}}|\psi_{4}\rangle_{A_{2}B_{2}}|b\rangle_{C_{1}}\right),\label{eq:the 5-qubit state}\end{equation}
where single qubit states $|a\rangle$ and $|b\rangle$ satisfy $\langle a|b\rangle=\delta_{a,b}$,
$|\psi_{i}\rangle\in\left\{ |\psi^{+}\rangle,|\psi^{-}\rangle,|\phi^{+}\rangle,|\phi^{-}\rangle:|\psi_{1}\rangle\neq|\psi_{3}\rangle,|\psi_{2}\rangle\neq|\psi_{4}\rangle\right\} $,
$|\psi^{\pm}\rangle=\frac{|00\rangle\pm|11\rangle}{\sqrt{2}},$ $|\phi^{\pm}\rangle=\frac{|01\rangle\pm|10\rangle}{\sqrt{2}}$.
This is clearly a special case of the more general state (\ref{eq:the 5-qubit state-2}),
and we may conclude that the 5-qubit states that are shown to be useful
for BCST and CBRSP are also useful for controlled bidirectional DSQC.
In our earlier work \cite{bi-directional-ourpaper}, we have already
shown that the total number of possible 5-qubit quantum states of
the form (\ref{eq:the 5-qubit state}) is infinite. Thus, there exist
infinitely many alternative 5-qubit quantum states that can be used
for implementation of bidirectional controlled DSQC.

\section{Qubit efficiency of the protocols\label{sec:Qubit-efficiency-of}}

The efficiency of a quantum cryptographic protocol is quantitatively
measured using two analogous but different parameters. The first one
is defined as \begin{equation}
\eta_{1}=\frac{c}{q},\label{eq:efficency  1}\end{equation}
where $c$ denotes the total number of transmitted classical bits
(message bits), and $q$ denotes the total number of qubits used \cite{the:C.-W.-Tsai}.
The limitation of this simple measure is that it does not include
the classical communication that is required for decoding of the information
in a DSQC protocol or CDSQC protocol. To circumvent this limitation
of the first measure, another quantitative measure \cite{defn  of qubit  efficiency}
is defined as \begin{equation}
\eta_{2}=\frac{c}{q+b},\label{eq:efficiency
    2}\end{equation}
where $b$ is the number of classical bits exchanged for decoding
of the message (classical communication used for checking of eavesdropping
is not counted). It is straightforward to visualize that $\eta_{1}=\eta_{2}$
for all QSDC and QSDC$^{{\rm GV}}$ protocols, but $\eta_{1}>\eta_{2}$
for all DSQC protocols {[}cf. \cite{With Anindita-pla} for a detail
discussion{]}. Hassanpour and Houshmand \cite{CQSDC-Hassanpour} compared
efficiency of their protocol of CDSQC with that of Gao et al. \cite{CQSDC-Gao}
and Dong et al. \cite{CQSDC-Dong} protocols using these quantitative
measures of efficiency. However, they have used a different notation
in which they referred to $\eta_{1}$ of the present paper as $\eta_{2}$
and vice versa, but that does not affect our analysis. 

It is important to note that the decoy qubits used for eavesdropping
check and classical communications involved for eavesdropping check
is not included in computation of qubit efficiency in Ref. \cite{CQSDC-Hassanpour}.
Remaining consistent with them, we may note that if any of the following
protocols $\left\{ {\rm CL,\, CL^{GV},\, DLL,\, DLL^{GV}}\right\} $
is used as a sub-protocol in our protocol of CDSQC, then each Bell
state can be used for the transmission of 2 bits of classical information,
implying that for the whole protocol $c=2n,$ and it requires $2n$
qubits, so $q=2n$. Further, disclosure of $\Pi_{n}$ requires $n$-bit
of classical information. Thus, $b=n$. This leads to $\eta_{2}=\frac{2n}{3n}=66.67\%$,\textbf{
}and $\eta_{1}=\frac{2n}{2n}=100\%.$ Clearly, this is more efficient
than the existing protocols. However, the efficiency calculation is
not appropriate, we should count the qubits used as decoy qubits.
In the above, $2n$ decoy qubits are used in total to check eavesdropping
during Charlie's transmission step. Another $n$ qubits are used for
eavesdropping check during Alice to Bob transmission. Thus, $q=2n+3n=5n$
and $\eta_{2}=\frac{2n}{6n}=33.33\%$ and $\eta_{1}=\frac{2n}{5n}=40\%.$

In Section 3.3 of Ref. \cite{CQSDC-Hassanpour}, some of the GHZ-like
states prepared by Alice are used for security check. Hassanpour and
Houshmand did not explicitly mention how many qubits are used for
security check. However, it is well known that to obtain the required
security, half of the transmitted qubits must be checked for eavesdropping.
Specifically, if $2x$ qubits (a random mix of message qubits and
decoy qubits) travel through a quantum channel accessible to Eve and
$x$ of them are tested for eavesdropping, then for any $\delta>0,$
the probability of obtaining less than $\delta n$ errors on the check
qubits (decoy qubits), and more than $(\delta+\epsilon)n$ errors
on the remaining $x$ qubits is asymptotically less than $\exp[-O(\epsilon^{2}x)]$
for large $x$ \cite{nielsen}. Thus, to obtain an unconditional security,
we always need to check half of the travel qubits for eavesdropping.
Therefore, to obtain 2 copies of GHZ-like state that are used in Eq.
(9) of Section 3.4 of Ref. \cite{CQSDC-Hassanpour}, Alice must start
the preparation phase (i.e., Section 3.2 of Ref. \cite{CQSDC-Hassanpour})
with 4 copies of GHZ-like state. Therefore, $q=12$ and consequently
corrected qubit efficiency of the protocol of Hassanpour and Houshmand
should be $\eta_{1}=\frac{2}{12}=16.66\%$, and $\eta_{2}=\frac{2}{12+3}=13.33\%$.
Clearly, the efficiency of the protocol reported here is 3 (2) times
more than that of the protocol of Hassanpour and Houshmand if we use
$\eta_{2}(\eta_{1})$ as the measure of efficiency. The CBDSQC scheme
described above has the same qubit efficiency as that of the unidirectional
scheme. This is so, because, in the bidirectional case all quantities
(e.g., $q,$ $m$ and $b$) just get doubled in comparison to their
values in unidirectional case. This linear change has no impact on
the efficiency. However, Alternative 3 described in Section \ref{sub:Alternative-3:-CDSQC}
will have different values of efficiency. To illustrate this fact,
we consider that the initial state used here is $n$ copies of GHZ-like
states. Therefore, Charlie keeps $n$-qubits (all the last qubits)
and sends the rest of the $2n$ qubits to Alice and Bob through the
quantum channel. For the secure transmission of those $2n$ qubits,
Charlie has to insert $2n$ decoy qubits, too. Thus, up to this step
we require $5n$ qubits. Using dense coding Alice can send $2n$ bits
of information, thus $c=2n$. However, during Alice to Bob communication,
Alice has to add $n$ decoy qubits that will lead to $q=6n$ and thus
$\eta_{1}=\frac{2}{6}=33.33\%$ and $\eta_{2}=\frac{2}{7}=28.57\%$.
However, efficiency of this scheme can asymptotically approach $40\%$.
To visualize this we may assume that the state prepared by Charlie
in Alternative 3 is $|{\rm Cat_{1}\rangle|0\rangle+|{\rm Cat_{2}\rangle|1\rangle}}$,
where ${\rm Cat_{i}}$ is a $2m$ qubit ${\rm Cat}$ state, where
maximal dense coding is possible. In this case, Charlie sends $m$
qubits to Alice and $m$ qubits to Bob with equal amount of decoy
qubits. Subsequently, after encoding $2n$ bits of information using
densecoding scheme, Alice would send Bob $m$ message qubits that
she received from Charlie and $m$ decoy qubits. Finally, Charlie
announces the outcome of his measurement in computational basis. This
implies $q=5m,\, c=2m$ and $b=1.$ Thus, $\eta_{2}=\frac{2m}{5m+1}$
and for $m\gg1,$ we obtain $\eta_{2}=\frac{2}{5}=40\%.$ 

\begin{table}
\begin{centering}
\begin{tabular}{|>{\centering}p{2.5in}|c|c|c|c|}
\hline 
Protocol & \multicolumn{2}{c|}{Without counting decoy qubits } & \multicolumn{2}{c|}{counting decoy qubits}\tabularnewline
\hline 
 & $\eta_{1}$ & $\eta_{2}$ & $\eta_{1}$ & $\eta_{2}$\tabularnewline
\hline 
HH \cite{CQSDC-Hassanpour} & $33.33\%$ & $22.22\%$ & $16.66\%$ & $13.33\%$\tabularnewline
\hline 
proposed CDSQC protocol (unidirectional using Bell states) & $100\%$ & $66.67\%$ & $40\%$ & $33.33\%$\tabularnewline
\hline 
proposed CDSQC protocol (bidirectional using Bell states) & $100\%$ & $66.67\%$ & $40\%$ & $33.33\%$\tabularnewline
\hline 
CDSQC using Alternative 3 (unidirectional with Bell states)  & $66.67\%$ & $50\%$ & $33.33\%$ & $28.57\%$\tabularnewline
\hline 
CDSQC using Alternative 3 (unidirectional with $(2m+1)$-qubit states
($m\gg1$))  & $100\%$ & $100\%$ & $40\%$ & $40\%$\tabularnewline
\hline
\end{tabular}
\par\end{centering}

\caption{\label{tab:Comparison-of-efficiency}Comparison of efficiency of HH
protocol \cite{CQSDC-Hassanpour} and the protocols proposed here.
In this table, we have not included the efficiency of CDSQC protocols
of Dong et al. \cite{CQSDC-Dong} and Gao et al. \cite{CQSDC-Gao}
as in Ref. \cite{CQSDC-Hassanpour} it is already established that
HH protocol is more efficient than Dong et al. and Gao et al. protocols.}
\end{table}

\section{Conclusions\label{sec:Conclusions}}

We conclude this paper by noting the following useful and important
observations:
\begin{enumerate}
\item It is shown that in the strict sense the protocol of Hassanpour and
Houshmand and the protocols designed here are actually protocols of
Controlled DSQC and not of Controlled QSDC as in these protocols,
and all the similar protocols, the controller has to disclose some
information without which the receiver will not be able to decode
the information encoded by the sender. 
\item In Ref. \cite{CQSDC-Hassanpour}, Hassanpour and Houshmand described
a protocol of CDSQC using entanglement swapping. For the purpose,
they have used two copies of GHZ-like states. Similar protocols can
be devised using a large number of entangled states of a generic form
described by Eq. (\ref{eq:state of interest}). Further, using block
streaming of qubits and PoP (as used in the first protocol described
here) the controller can securely distribute quantum states of the
form (\ref{eq:state of interest}) to the receiver and sender, and
subsequently they can use QSDC/DSQC protocol described in our earlier
work \cite{dsqc-ent swap}. This point is already elaborated in Section
\ref{sec:Controlled-secure-direct}. In Sections \ref{sec:Controlled-secure-direct}
and \ref{sec:Controlled-bidirectional-secure}, we have also provided
a large number of alternative approaches and alternative quantum states
that lead to CDSQC. Thus, this paper provides many alternatives for
experimental realizations of CDSQC.
\item Here, it is established that the efficiency reported in the efficient
CDSQC protocol of Hassanpour and Houshmand can be considerably increased
by appropriately using block streaming and PoP. The point is clearly
illustrated in the previous section and in Table \ref{tab:Comparison-of-efficiency}. 
\item The fact that the efficiency of entangled state based protocol is
less compared to the proposed protocols is not surprising as in the
context of conventional QSDC and DSQC protocols designed by us \cite{With Anindita-pla,With chitra-ijtp,With chitra IJQI},
we have already seen that using block streaming and PoP we can design
maximally efficient protocols of QSDC and DSQC, but the maximal efficiency
was not achieved when we used entanglement swapping for the same purpose
{[}cf. our earlier work \cite{dsqc-ent swap}{]}.
\item In any protocol of CDSQC the receiver and sender need to be semi-honest
\cite{switch}. Otherwise, it will always be possible for Alice and
Bob to avoid the control of Charlie and share the secret information
(or quantum state) by creating a quantum channel of their own. For
example, dishonest Alice and Bob may decide to use CL or DLL protocol
in their original form and completely ignore Charlie. However, such
situation will not arise if we consider Alice and Bob as semi-honest
as the semi-honest users follow the protocol, but tries to cheat the
controller remaining within the protocol. This very important point
was not realized by Hassanpour and Houshmand, and consequently they
restricted their discussion to the external attacks (attacks of Eve).
However, for a protocol of CDSQC, it is important to show that the
protocol is secure from internal attacks of semi-honest receivers
and senders. To do so, it is desirable (but not essential) that the
controller prepares the state (quantum channel). Otherwise, Alice
can supply a separable qubit to Charlie and thus get rid of his control.
This desirable condition is not followed in HH protocol. Consequently,
Alice can always cheat Charlie, but the proposed protocol is free
from such an internal attack.
\item All the protocols of CDSQC that are introduced until now \cite{CQSDC-Hassanpour,CQSDC-Gao,CQSDC-Dong}
are conjugate-coding based. In these protocols, security arises from
the use of two or more mutually unbiased bases (MUBs). In contrast
to them, here we have shown that if we use GV subroutine for eavesdropping
checking in all the steps of the proposed protocols then we obtain
orthogonal-state-based protocols of CDSQC. This is fundamentally important
as it establishes that protocols of CDSQC can be achieved without
using conjugate coding (non-commutativity).
\end{enumerate}
In brief, in this paper, we have established that it is possible to
construct a large number of alternative protocols of unidirectional
and bidirectional efficient controlled secure direct quantum communication
by using various quantum states. Thus, the present theoretical work
provides a large number of alternatives for the experimental realization
of a CDSQC or CBDSQC scheme. Keeping this fact in mind, we conclude
the paper with an expectation that the protocols proposed in this
paper will be experimentally realized in near future.

\textbf{Acknowledgments:} AP thanks Department of Science and Technology
(DST), India for the support provided through the DST Project No.
SR/S2/LOP-0012/2010. He also thanks K. Thapliyal and C. Shukla for
carefully reading the manuscript.

\end{document}